\documentclass[runningheads]{llncs}

\usepackage[T1]{fontenc}
\usepackage{graphicx}

\usepackage{amsmath,amssymb,amsfonts}
\usepackage[inline, shortlabels]{enumitem}
\usepackage{tabularx}
\usepackage{caption}
\usepackage{listings}
\usepackage[english]{babel}
\usepackage{ragged2e}
\usepackage[hyphens]{url}
\usepackage{hyperref}
\usepackage{xurl}
\usepackage{pifont}
\usepackage{footmisc}
\usepackage{multirow}
\usepackage{enumitem}
\usepackage{algorithm2e}
\usepackage{float}
\usepackage{listings}
\usepackage{xcolor}

\definecolor{codegreen}{rgb}{0,0.6,0}
\definecolor{codegray}{rgb}{0.5,0.5,0.5}
\definecolor{codepurple}{rgb}{0.58,0,0.82}
\definecolor{backcolour}{rgb}{0.95,0.95,0.92}
 
\lstdefinestyle{mystyle}{
    backgroundcolor=\color{backcolour},   
    commentstyle=\color{codegreen},
    keywordstyle=\color{magenta},
    numberstyle=\tiny\color{codegray},
    stringstyle=\color{codepurple},
    basicstyle=\footnotesize,
    breakatwhitespace=false,         
    breaklines=true,                 
    captionpos=b,                    
    keepspaces=true,                 
    numbers=left,                    
    numbersep=5pt,                  
    showspaces=false,                
    showstringspaces=false,
    showtabs=false,                  
    tabsize=2
}
 
\usepackage{physics}
\usepackage{amsmath}
\usepackage{tikz}
\usepackage{mathdots}
\usepackage{yhmath}
\usepackage{cancel}
\usepackage{color}
\usepackage{siunitx}
\usepackage{array}
\usepackage{multirow}
\usepackage{amssymb}
\usepackage{gensymb}
\usepackage{tabularx}
\usepackage{extarrows}
\usepackage{booktabs}
\usetikzlibrary{fadings}
\usetikzlibrary{patterns}
\usetikzlibrary{shadows.blur}
\usetikzlibrary{shapes}

\usepackage{pdfpages}
\usepackage{booktabs}
\usepackage{csquotes}
\usepackage{lipsum}  
\usepackage{arydshln}
\usepackage{smartdiagram}
\usepackage[inkscapeformat=png]{svg}
\usepackage{textcomp}
\usepackage{tabularray}\UseTblrLibrary{varwidth}
\usepackage{xcolor}
\def\BibTeX{{\rm B\kern-.05em{\sc i\kern-.025em b}\kern-.08em
    T\kern-.1667em\lower.7ex\hbox{E}\kern-.125emX}}
\usepackage{cite}
\usepackage{amsmath}
\newcommand{\probP}{\text{I\kern-0.15em P}}
\usepackage{etoolbox}
\patchcmd{\thebibliography}{\section*{\refname}}{}{}{}

\newcounter{relation}
\renewcommand{\therelation}{\arabic{relation}}

\newcounter{proof}
\renewcommand{\theproof}{\arabic{proof}}

\begin{document}
\title{A MARL-based Approach for Easing MAS Organization Engineering}
%
%
\author{Julien Soulé\inst{1} \and
Jean-Paul Jamont\inst{1} \and
Michel Occello\inst{1} \and
Louis-Marie Traonouez\inst{2} \and
Paul Théron\inst{3}}
\authorrunning{J. Soulé et al.}
%
\institute{Univ. Grenoble Alpes, Grenoble INP, LCIS, 26000, Valence, France
    \email{\{julien.soule, jean-paul.jamont, michel.occello\}@lcis.grenoble-inp.fr}
    \and
    Thales Land and Air Systems, BL IAS, Rennes, France
    \email{louis-marie.traonouez@thalesgroup.com}
    \and
    AICA IWG, La Guillermie, France \\
    \email{paul.theron@orange.fr}
}

\maketitle              


\begin{abstract}

    Multi-Agent Systems (MAS) have been successfully applied in industry for their ability to address complex, distributed problems, especially in IoT-based systems.
    Their efficiency in achieving given objectives and meeting design requirements is strongly dependent on the MAS organization during the engineering process of an application-specific MAS. To design a MAS that can achieve given goals, available methods rely on the designer's knowledge of the deployment environment.
    However, high complexity and low readability in some deployment environments make the application of these methods to be costly or raise safety concerns.
    In order to ease the MAS organization design regarding those concerns, we introduce an original Assisted MAS Organization Engineering Approach (AOMEA). AOMEA relies on combining a Multi-Agent Reinforcement Learning (MARL) process with an organizational model to suggest relevant organizational specifications to help in MAS engineering.

\end{abstract}

\section{Introduction}


MAS have drawn significant interest in the industrial field due to their ability to address complex, distributed problems~\cite{Raileanu2023}.
That paradigm enables decomposing a complex task into missions that are delegated to autonomous agents that achieve them through cooperation mechanisms. Most notably, they provide models and approaches to handle conflicting goals, parallel computation, system robustness, and scalability.
In MAS, the organization is a fundamental concept that has an impact on how agents coordinate their activities to collaboratively achieve a common goal~\cite{Hubner2007}.
Organizational aspects address the challenge of MAS design in dynamic and uncertain environments, where runtime behavior needs to be flexible~\cite{Kathleen2020}. Organization in MAS design is a central concept in methodologies and frameworks enabling the engineering of application-specific MAS~\cite{Bakliwal2018}.

MAS design/development methods have often been proposed jointly with organizational models to help designers find suitable organizational specifications enabling a MAS to reach a goal efficiently. Methods such as GAIA~\cite{Wooldridge2000,Cernuzzi2014}, ADELFE~\cite{Mefteh2015}, or DIAMOND~\cite{Jamont2015}, KB-ORG~\cite{Sims2008} provide protocols that rely on the designer's experience to hand-craft the agent's rules (also called \textbf{policies}) leveraging \textbf{self/re-organization} mechanisms to adapt the MAS on the deployment environment.
These aforementioned methods are commonly applied through simulations for they enable a safe monitoring framework for the design process and assessment. A MAS developed in simulated environments with high fidelity to the target system is expected to be transferred to the target system to perform adequately~\cite{Schon2021}.

The designer defines the agents' policies in various ways ranging from the agent's individual point of view to the global organization point of view. A properly designed MAS is expected to show emerging or chosen organizations enabling reaching a goal~\cite{Picard2009}. That design approach often takes place as an iterative process proceeding by trial and error. Yet, it shows the following limitations:
\begin{enumerate*}[label=\roman*),itemjoin={;\quad}]
    \item It requires sufficiently experienced designers
    \item It may be costly to converge towards a sufficiently estimated successful MAS
    \item It gets difficult to apply for complex and highly dimensional target deployment environments.
\end{enumerate*}
For instance, research in Autonomous Intelligent Cyberdefense Agents~\cite{Kott2023} (AICA) aims to develop cooperative Cyberdefense agents deployed in highly complex computer networks. The development of an AICA faces the lack of visual and intuitive comprehension of the networked environments such as company networks.


Even though some methods may automate some parts of the MAS organization design such as KB-ORG~\cite{Sims2008}, they still require some knowledge and manual interactions to guide the designing process. Indeed, there is a need for
\begin{enumerate*}[label=\roman*),itemjoin={; and \ }]
    \item Finding automatically suited agents' policies satisfying design constraints
    \item Making explicit the organizational mechanisms that emerge from trained agents for the design process.
\end{enumerate*}


To address these issues, we introduce AMOEA, a MAS design approach whose underlying idea is to link a given MARL process with an organizational model that links the on-training agents' policies with explicit organizational specifications. It can be viewed as a tool for engineering to automatically generate relevant exploitable organizational specifications only regarding the performance in achieving the given goal and the design constraints. For the designer, the obtained organizational specifications are insights into the organizational mechanisms to set up for developing a MAS that meets performance requirements.



Section II starts by introducing the theoretical background of AOMEA and focuses on the fundamental concepts we used for the organizational models and MARL.
In section III, we present AOMEA from the approach to the implemented tool. We assessed AOMEA in four simulated environments and discussed the obtained results in section IV. Finally, section V concludes on the AOMEA's viability and highlights limitations to overcome and future works as well.


\section{Theoretical background}





In this section, we present the basics of the $\mathcal{M}OISE^+$ organizational model and the MARL basics on which our contribution is built.

\subsection{Multi-Agent Systems context}


An agent is an entity immersed in an environment that perceives observation and makes a decision to act autonomously in the environment to achieve the objectives assigned to it.
Agent types include event-driven reactive to deal with uncertainties in an environment or cognitive proactive agents that leverage interactions with other agents. A MAS is a set of agents in a shared environment where each agent has only a local perception. These agents are to be endowed with self/re-organizing capabilities that allow them to adaptively modify their organizational structure according to their environment.

A MAS is strongly linked to the organization entity (we simply call \textbf{organization}) we consider it to always exist through the running agents' interactions even though it may be implicit.
%
%
An \textbf{organizational model} specifies (at least partially) the organization whether it is used as a medium to describe an explicit known organization in a top-down way, or describing an implicit organization in a bottom-up way. An example of organizational model is the \emph{Agent/Group/Role} (AGR) model~\cite{Ferber2004}. We refer to the \textbf{organizational specifications}, the components used in an organizational model to characterize the organization. $\mathcal{M}OISE^+$ is an organizational model with which it is possible to link agents' policies to organizational specifications. It takes into account the social aspects between agents explicitly whereas \emph{AGR} focuses on the integration of standards oriented towards design. $\mathcal{M}OISE^+$~\cite{Hubner2007} considers three types of specifications:

The \textbf{structural specifications} describe the means agents can leverage to achieve a goal. It comprises the set of \emph{roles}, sub-groups, intra-group and inter-group \emph{links}, intra-group and inter-group \emph{compatibilities}, and the role and sub-group \emph{cardinalities}.
A \emph{link} indicates whether two roles are related because of acquaintance, communication, or authority ties. A \emph{compatibility} indicates whether two roles can be adopted by the same agent. Role and sub-group \emph{cardinalities} respectively refer to the minimal and maximal number of roles and sub-groups.

The \textbf{functional specifications} describe the way to achieve a goal. It comprises \emph{social schemes} and \emph{preference order}. A \emph{social scheme} is described by global goals, mission labels with plans, and the cardinality of agents committed to a mission. A \emph{preference order} means an agent has a social preference to commit to a specific mission among several possible ones.

The \textbf{deontic specifications} enable linking functional and structural specifications through a set of \emph{permissions} and obligations. A \emph{permission} means an agent playing role $\rho_a$ is permitted to commit to mission $m$ for a given time constraint $tc$. Similarly, an \emph{obligation} means an agent playing role $\rho_a$ has to commit to mission $m$ for a given time constraint $tc$. A time constraint $tc $ specifies a set of periods determining whether a permission or an obligation is valid.


\subsection{MARL basics}

Reinforcement learning is a machine learning paradigm where agents learn to make decisions by interacting with an environment. The goal is for the agent to maximize a cumulative reward signal over time through a trial-and-error process.
MARL extends this concept to multiple agents that learn while considering the actions of other agents pushing agents to rely on cooperation mechanisms.

MARL enables automatically converging towards agents’ policies that enable reaching the given goal. Yet, unlike human-based design, the trained agents' logic is explicitly specified from a collective point of view. Few works attempt to address that issue and few are oriented for methodological purposes.
Kazhdan et. al.~\cite{Kazhdan2020} proposed means to extract symbolic models from MARL systems that improve the interpretability of MARL systems.
Wang et. al.~\cite{Wang2020} introduced a role-oriented MARL approach where roles are emergent, and agents with similar roles tend to share their learning and specialize in certain sub-tasks.
Tosic et. al~\cite{Tosic2010} proposed a framework for addressing coordination in collaborative MAS relying on the communication capabilities of multi-agent systems.
Zheng et. al.~\cite{Zheng2018} presented a platform for MARL that aims to facilitate research on artificial collective intelligence by providing a comprehensive set of baselines and evaluation metrics to benchmark the performance of MARL algorithms.

Markovian models are required to model the environment and apply MARL techniques. As a commonly used, Decentralized Dec-POMDP~\cite{Oliehoek2016} considers multiple agents in a similar MAS fashion. It relies on stochastic processes to model the uncertainty of the environment for the changes induced by the actions, the received observations, and the communications as well. Its reward function is common to agents which fosters training for collaborative oriented actions~\cite{Beynier2013}. Formally, a Dec-POMDP is a 7-tuple $(S,\{A_i\},T,R,\{\Omega_i\},O,\gamma)$ , where: $S = \{s_1, ..s_{|S|}\}$: The set of the possible states; $A_{i} = \{a_{1}^{i},..,a_{|A_{i}|}^{i}\}$: The set of the possible actions for agent $i$; $T$ so that $T(s,a,s') = \probP{(s'|s,a)}$ : The set of conditional transition probabilities between states; $R: S \times A \times S \rightarrow \mathbb{R}$: The reward function; $\Omega_{i} = \{o_{1}^{i},..,o_{|\Omega_{i}|}^{i}\}$: The set of observations for agent $ag_i$; $O$ so that $O(s',a,o) = \probP{(o|s',a)}$ : The set of conditional observation probabilities; $\gamma \in [0,1]$, the discount factor.

We refer to \textbf{solving} the Dec-POMDP for the team $t$ as finding a joint policy $\pi_{joint,i} \in \Pi_{joint}$ that maximizes the expected cumulative reward over a finite horizon.
We refer to \textbf{sub-optimally solving} the Dec-POMDP at $s$ expectancy as finding the joint policies $\pi_{joint,i} \in \Pi_{joint}$ that gets the expected cumulative reward over a finite horizon at least at $s \in \mathbb{R}$.

\section{AOMEA approach}



\subsection{Overview}

\begin{figure}[h!]
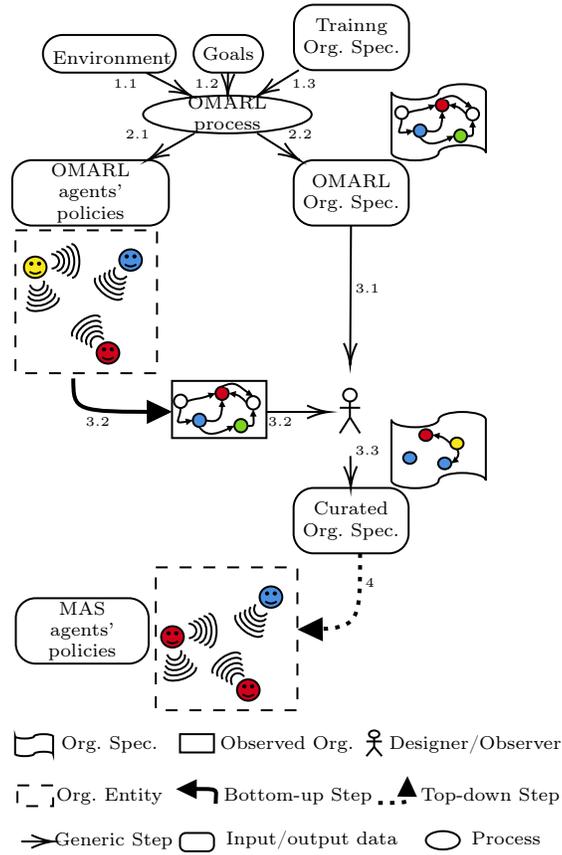

    \centering
    \include{figures/approach}
    \caption{A summary view of our approach to MAS design}
    \label{fig:design_approach}
\end{figure}

We introduce AOMEA as an approach for MAS design that automates the preliminary design of a MAS according to some design constraints. Organizational specifications obtained after training allow the development of a curated MAS.
The underlying idea of our approach is to consider that a joint-policy or joint-history can be described in terms of organizational specifications, at least partially.
We refer to that broad approach as \textquote{Organization oriented MARL} (OMARL).
%
%
AOMEA consists of 4 sequential phases: modeling, solving, analyzing, and developing (respectively $1.x$, $2.x$, $3.x$, $4.x$ in arrow labels in \autoref{fig:design_approach}).

\textbf{Phase 1: Modeling} \quad In that phase, the designer has to manually develop a simulation of the target environment ($1.1$ in \autoref{fig:design_approach}) where agents must cooperate to achieve the designer's goal efficiently ($1.2$ in \autoref{fig:design_approach}) with the help of quantitative feedback. When developing the simulated environment, the designer can link parts of an agent's policy (as observations-actions couples) with known organizational specifications of any chosen organizational model.
For instance, in \textquote{leader-follower} organizations, the actions that send orders to other follower agents, are characteristics of leader agents.
Optionally, the designer may also want to restrict the set of possible policies agents can explore regarding given organizational specifications as constraints to meet design requirements or to help agents converge as well ($1.3$ in \autoref{fig:design_approach}).

\textbf{Phase 2: Solving} \quad In that phase, relying on the established relations between observation-action couples and organizational specifications, a MARL algorithm is used jointly with the chosen MAS organizational model through an OMARL process. It automatically enables finding optimal policies satisfying the given design organizational specifications ($2.1$ in \autoref{fig:design_approach}) that lead to the best expected cumulative reward; and getting the associated organizational specifications ($2.2$ in \autoref{fig:design_approach}). For instance, when training agents regarding the \textquote{leader-follower} organization, some agents may be forbidden to send orders while some other may be forced to. After training, the OMARL process characterizes emergent roles, links between roles, or sub-goals organized in plans to reach the goal.

\textbf{Phase 3: Analyzing} \quad In that phase, the designer observes the trained agents' policies ($3.2$ in \autoref{fig:design_approach}) and takes into account the inferred associated organizational specifications ($3.1$ in \autoref{fig:design_approach}) to understand how these agents can reach the goal. In light of these raw results, the designer can extract valuable design patterns from noisy or useless agents' decisions. The interest is to provide at least some indications of the organizational specifications capable of achieving the goal and to satisfy the design constraints. We refer to these valuable indications as curated organizational specifications ($3.3$ in \autoref{fig:design_approach}). For instance, after having trained several agents in a \textquote{predator-prey} environment, it is possible to analyze that a \textquote{leader} predator with \textquote{follower} predators, appears to be more efficient for catching prey.

\textbf{Phase 4: Developing} \quad In that phase, the designer takes into account the curated organizational specifications as a blueprint for implementing a MAS. From that point, a regular MAS development with one of the available methods that is used jointly with the chosen organizational model can be applied. Unlike the trained agents which may cause unexpected behavior, manually implemented agents enable giving safety guarantees required for sensitive environments. Finally, implemented agents are launched in simulations to assess whether the implemented MAS can effectively achieve the goal.

\subsection{Theoretical core}

\begin{figure}[h!]
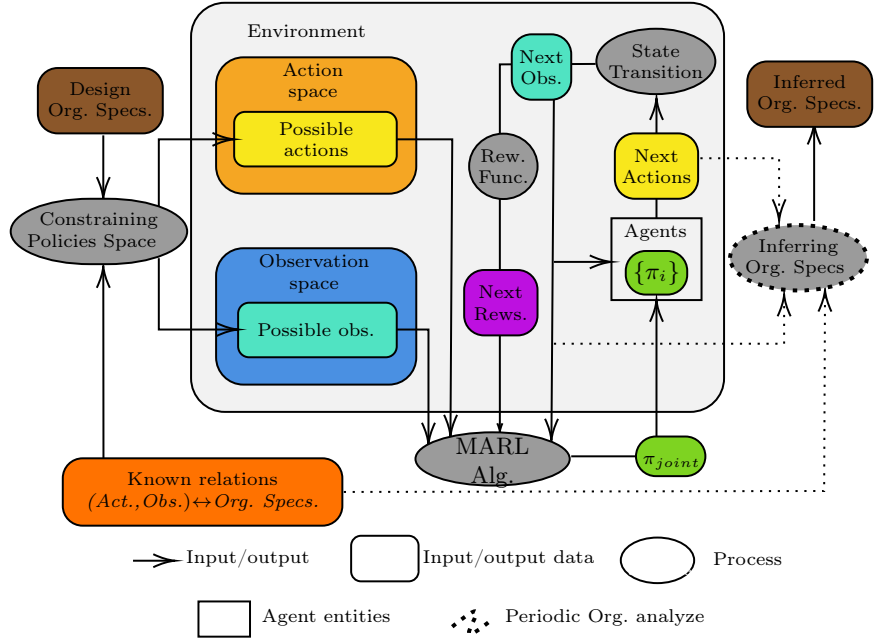

    \centering
    \include{figures/prahom_process}
    \caption{A summary view of the PRAHOM process}
    \label{fig:prahom_process}
\end{figure}

To implement an OMARL process, we propose the \emph{Partial Relations with Agent History and Organization Model} algorithm (PRAHOM) to link agents' policies and their training to an organizational model.
It is a synthesis of two processes that fall into the OMARL purposes. The first process gets the specifications from the agents' policies, and the second process gets the joint-policies satisfying the given design specifications. An illustrative view of \emph{PRAHOM} is given in \autoref{fig:prahom_process}.
Here we just present the underlying idea at a high-level description for these two processes to avoid unnecessary formalism. More information on the use and implementation of \emph{PRAHOM} can be found in \autoref{PettingZoo-wrapper}.

\paragraph{\textbf{Inferring Organizational Specifications}}

Rather than using joint-policies directly, we use the joint-histories since they may be built with observed resulting actions when observations are received during a series of test episodes. Indeed, for a given policy $\pi \in \Pi$, the associated history is by definition $h \in h_{joint} = \langle(\omega_k,a_k) | k \in \mathbb{N}\rangle$ and the $(\omega_k,a_k) \in \pi$.
Then, due to the difficulty of inferring information related to organizational specifications, it is possible to associate each observation or action with organization specifications as a \textquote{many to many} relation. It sets up a first frame for identifying organizational specifications in histories. We address that problem in the remainder of this section.

One can define some relations between $\mathcal{M}OISE^+$ specifications and joint-histories. Their premises come from noticing some specifications in the $\mathcal{M}OISE^+$ organizational model can be mapped to subsets of actions from a single suboptimal joint-policy.
From these relations, it is possible to use empirical or statistical approaches to infer organizational specifications out of joint-histories. Below we informally describe key points for understanding that process.

As we have only one group, we do not consider the inter-links and inter-compatibilities. Additionally, as a simplification, we consider only one social scheme.
First, we look at the individual level by trying to figure out the roles, links, sub-groups, individual goals, missions, and plans played by agents by sampling history subsequences $h \in H$ and comparing with known history subsequences whose we know the associated role via the established relations.

After analyzing several joint-policies, we try to reinforce a global view of the goals, missions, plans, and the knowledge of the mission to the goal; with the partially inferred information at the individual level.
In the end, our process tries to synthesize the knowledge inferred until having a better view of the agent cardinality per sub-group, the agent cardinality for each mission, the role cardinality, the compatibilities between roles, the permissions, and obligations.

\paragraph{\textbf{Constraining Policies Space}}

We consider a given MARL algorithm that iteratively converges towards a joint-policy so that each agent's policy is updated at each step until a finite horizon.
We favored the Proximal Policy Optimization for its proven effectiveness in cooperative multi-agent environments without the need for domain-specific algorithmic modifications or architectures~\cite{Yu2022}.
To constrain the possible joint-policies to the ones satisfying the design organizational specifications $os_{init}$, we propose to constrain the action and observation sets for each agent according to $os_{init}$ at each step. Ultimately, it constrains an agent to a role by forbidding actions related to other roles.

First, we use the established relations between organizational specifications and action-observation couples, to determine the authorized or forbidden actions playable by agents at each step.
Then, it first computes the authorized actions set $A_{step}$ according to the current history $h_{joint,i}$. Then, an action is chosen among authorized actions. That action $a_{step} \in A_{step}$ is added in history to be used for updating the agent's policy in the next step. Then, the MARL algorithm updates the joint-policy hence the agents' policies with the current action and observation.
Finally, an analysis of the current suboptimal joint-policy $\pi_{joint}$ satisfying $os_{init}$ is triggered periodically. It enables iteratively improving the efficiency of joint-policies and the accuracy of the inferred organizational specifications.
We can note the restriction implied by $os_{init}$ in the possible joint-policies might prevent the MARL algorithm from finding a joint-policy that satisfies the minimal expected cumulative reward defined by the designer.

\subsection{Engineering tool}

PettingZoo is a library that offers a standardized API that simplifies the development of environments with agents and facilitates the application of MARL algorithms.
We developed \emph{PRAHOM PettingZoo Wrapper}\label{PettingZoo-wrapper}, a tool to help automate the setting up of \emph{PRAHOM} for a given PettingZoo environment.
It is a \emph{PoC} linking joint-histories with $\mathcal{M}OISE^+$ specifications to provide functions to infer raw organizational specifications or constrain the training. 
\begin{lstlisting}[language=Python, caption=PRAHOM PettingZoo Wrapper basic use, label={lst:wrapper_basic_use}]
from omarl_experiments import prahom_wrapper
env=PettingZoo_env.parallel_env(render_mode="human")
specs_to_hist={"structural_specifications":{"roles":{"follower":{"23":41,"14":[74,0]}}...},"functional_specifications":{"links":{"(leader,follower,aut)":".*14.*?89"}...}...}
policy_specs_constr={"agent_0":{"structural_specifications":"roles":["follower"]}}
env=prahom_wrapper(env,action_to_specs,training_specs)
env.train("default_PPO")
trained_specs,agent_to_specs=env.prahom_specs()
\end{lstlisting}
In \autoref{lst:wrapper_basic_use}, we detailed a basic use of the wrapper to augment a PettingZoo environment (l. 5) with known relations between histories and organizational specifications (l. 3) and the design constraints agents are to satisfy (l. 4).
During training, \textquote{"agent\_0"} is constrained to role \textquote{follower} so all of its actions must be chosen regarding the relations between organizational specifications and expected histories (or shortened histories expressions).
After training, \emph{PRAHOM PettingZoo Wrapper} infers organizational specifications from joint-histories in 5 episodes, and the agents' instantiation for each one (l. 7).

This process first uses known relations between histories and organizational specifications (l. 3): an agent's history that contains the observation $14$ (\textquote{order received}) and the action $74$ (\textquote{apply order}) or $0$ (\textquote{do nothing}), is a \textquote{follower}. Similarly, history can be linked to organizational specifications with regular expressions such as for \emph{links} (l.3).
Then, it generalizes several joint-histories as new organizational specifications relying on their respective general definition.
For instance, a role is thought to be inferred by measuring similarity between histories in several ways: sequence clustering (with a dendrogram); K-nearest neighbors (with PCA of histories); statistical analysis (specially action frequency in various visualizations); etc.
Techniques are also used to infer goals: frequency analysis of common observations of agents with the same role; analysis of threshold states triggering an improvement of the reward (with a state transition graph).

From the roles and goals obtained, an empiric approach allows inferring the other organizational specifications such as compatibilities, permissions and obligations.
Due to that empiric approach and the specific scope of the techniques, results may be incomplete or noisy. Yet, since results are compliant with $\mathcal{M}OISE^+$, it is possible to use them in MAS design methods in light of the identified organizational specifications.

\section{Evaluation in cooperative game environments}

\begin{figure}[H]
    \centering
    \includegraphics[width=0.8\linewidth]{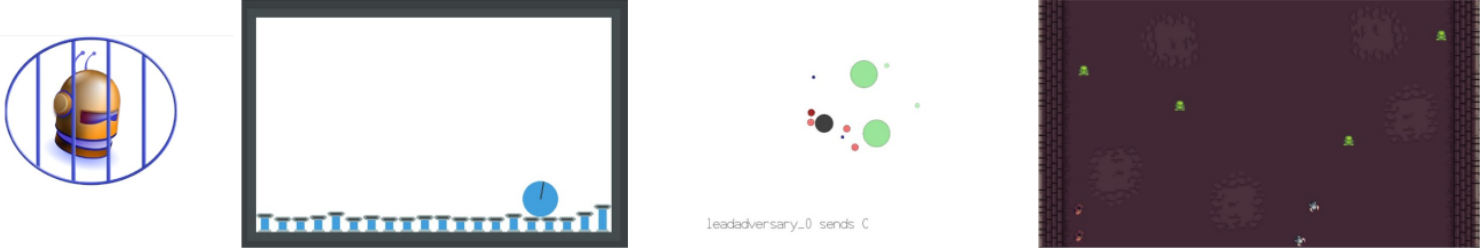}
    \caption{Overview of the selected environments: CYB, PBL, PPY, and KAZ}
    \label{fig:simulated_environments}
\end{figure}

In order to assess AOMEA, we considered using \emph{PRAHOM} in available simulated environments made up of agents that have to achieve a goal with the best performance through various collective strategies whose some can be easily understood (presented in \autoref{fig:simulated_environments}).
We selected three Atari-like environments for their visual rendering is a convenient way to assess the results with manual observations\footnotemark[1].
We also considered a Cyberdefense environment as a first attempt to apply \emph{PRAHOM} in a non-visual Cyberdefense environment:

\footnotetext[1]{Additional explanation and the examples discussed using \emph{PRAHOM PettingZoo wrapper} are available at \url{https://github.com/julien6/omarl_experiments?tab=readme-ov-file\#tutorial-predator-prey-with-communication}}

\begin{itemize}
    \item \textquote{Drone swarm - 3rd CAGE Challenge}~\cite{cage_challenge_3_announcement} (CYB) consists of cyberdender agents deployed on networked drones fighting against maliciously deployed malware programs. We may expect agents to \allowbreak isolate compromised drones;
    \item \textquote{Pistonball} (PBL)~\cite{Terry2021} consists of a series of pistons to bring a ball from right to left side hence requiring neighbors' representation;
    \item \textquote{Predator-prey with communication}~\cite{Lowe2017} (PPY) consists of predators monitored by a leader to catch faster prey hence requiring hunting strategies;
    \item \textquote{Knights Archers Zombies}~\cite{Terry2021} (KAZ) consists in knights and archers learning how to kill zombies hence requiring efficient agent spatial positioning.
\end{itemize}
\noindent We applied AOMEA in three cases:
\begin{itemize}
    \item No organizational specifications (NTS): agents have to learn the most efficient collective strategies without any constraints or indications.
    \item Partially constraining organizational specifications (PTS): some constraints or indications are given to help converge faster or meet requirements.
    \item Fully constraining organizational specifications (FTS): manually crafted joint-policies are given for they are a reference regarding learned joint-policies.
\end{itemize}

\noindent Here, we do not present the details of the constraints that were given in NTS and FTS (available in Git repository\footnotemark[1]).
\begin{figure}[h!]
    \centering
    \includegraphics[width=0.8\textwidth]{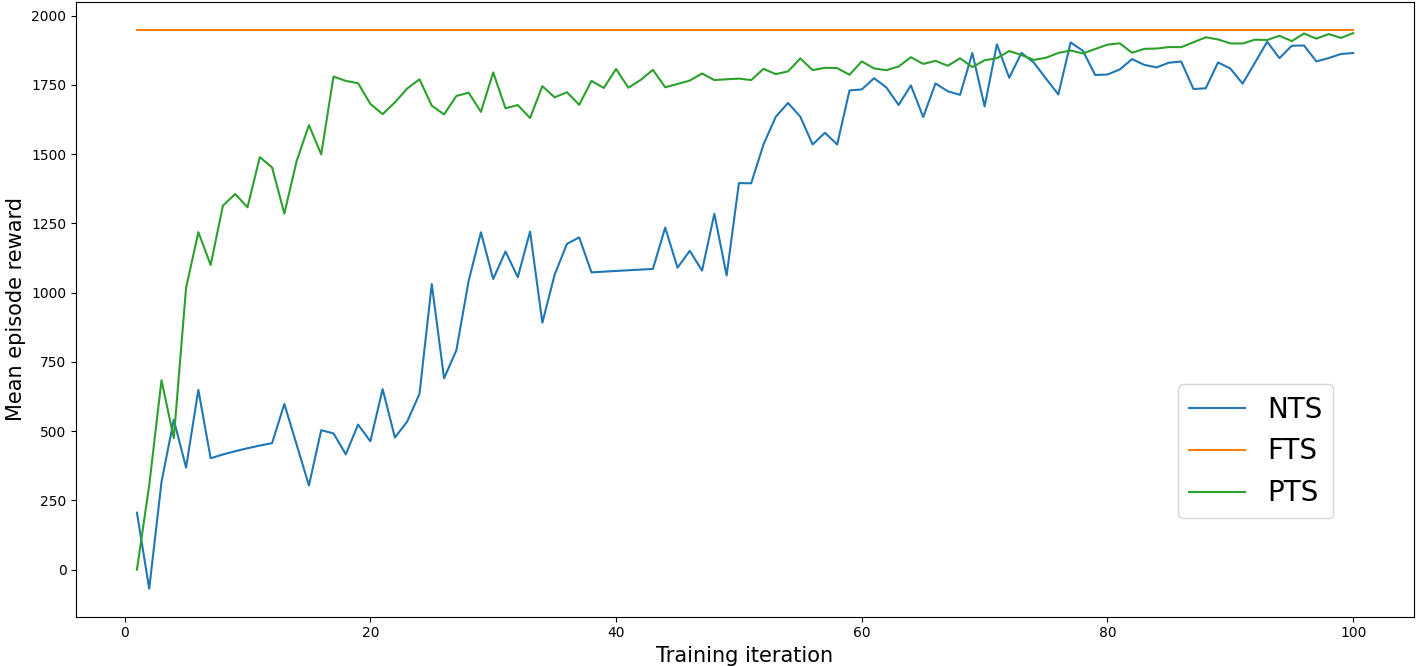}
    \caption{Average reward for each iteration in the PBL environment for the NTS, PTS, and FTS cases}
    \label{fig:prahom_learning_curve}
\end{figure}
\begin{figure}[h!]
    \centering
    \includegraphics[width=0.8\textwidth]{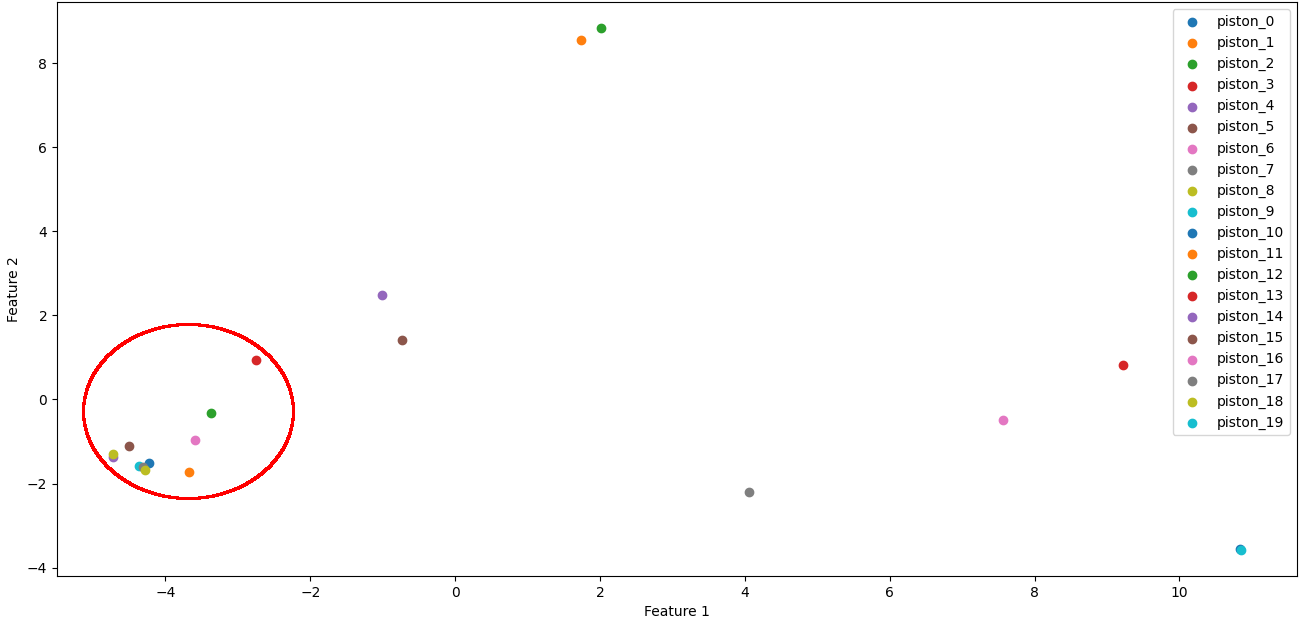}
    \caption{PCA of the trained agents' histories in the PBL environment}
    \label{fig:prahom_pca_analysis}
\end{figure}
We evaluate the impact of \emph{PRAHOM} on the following criteria: convergence time ratios between PTS, NTS, and FTS for reaching a threshold cumulative reward. Performance stability shows how the trained agents can achieve the goal generally by assessing several environments generated with different parameters. Results are presented in Table~\ref{tab:training_AOMEA_results}.
\begin{table}[t!]

    \centering

    \caption{View of the AOMEA approach impact during training in the PTS case}

    \begin{tblr}{colspec={llll},row{1}={rowsep=1mm,m},row{2-Z}={rowsep=-1mm,m},measure=vbox,stretch=-10}

        \textbf{Environment} & \textbf{NTS/PTS} & \textbf{PTS/FTS} & \textbf{Perf. stability \\ (avg. / max)} \\

        \hline

        { PBL }
        & { 4.7 }
        & { 1.3 }
        & { 0.9 } \\

        \hline[dashed]

        { PPY }
        & { 6.3 }
        & { 2.2 }
        & { 0.78 } \\

        \hline[dashed]

        { KAZ }
        & { 4.0 }
        & { 1.1 }
        & { 0.71 } \\

        \hline[dashed]

        { CYB }
        & { 12.0 }
        & { 3.3 }
        & { 0.36 } \\

    \end{tblr}

    \label{tab:training_AOMEA_results}

\end{table}

As a general observation, we can notice convergence time is longer for NTS than for PTS which is also longer than for FTS. As expected, the search space is decreasing, hence a shorter convergence time. For instance, we noticed a faster convergence to a sub-optimal solution in the PBL environment by providing organizational specifications as presented in \autoref{fig:prahom_learning_curve}. Although PTS converges faster than NTS to a comparable cumulative reward, NTS may outperform PTS because trained agents' policies are hand-tailored to solve the problem much more finely than the designer's organizational specifications can do. Low-performance stability in the more complex CYB environment indicates that the trained agents have difficulty finding general strategies compared to the agents in the other environments.

We also took into account the following criteria after training: roles, links, and global performance. A qualitative analysis is presented in Table~\ref{tab:trained_AOMEA_results}
\begin{table}[t!]

    \centering

    \caption{Qualitative analysis of the inferred organizational specifications after training in the NTS case}

    \begin{tblr}{colspec={llll},row{1}={rowsep=1mm,m},row{2-Z}={rowsep=0.5mm,m},measure=vbox,stretch=-10}

        \textbf{Environment} & \textbf{Roles emergence} & \textbf{Links emergence} & \textbf{Global \\ performance} \\

        \hline

        { PBL }
        & { Clear emerging \\ roles}
        & { Local \\ representation }
        & { Close \\ to optimal } \\

        \hline[dashed]

        { PPY }
        & { Inherently \\ differentiated }
        & { Rare strategies}
        & { Highly variable } \\

        \hline[dashed]

        { KAZ }
        & { Inherently \\ differentiated }
        & { Local \\ representation }
        & { Highly variable } \\

        \hline[dashed]

        { CYB }
        & { No clear \\ emerging roles }
        & { Some apparent \\ strategies }
        & { Quite good } \\

    \end{tblr}

    \label{tab:trained_AOMEA_results}

\end{table}

%
%
For the PBL environment, we can notice roles being equivalent for agents are expected to act the same. Indeed, trained agents' histories are close hence showing a common emerging role. We generate the PCA presented in \autoref{fig:prahom_pca_analysis} by expressing agents' histories as vectors containing the observation-action couples. We can notice most agents’ histories are in the left bottom zone (circled in red). It shows most pistons seem to act similarly as expected. We observe no organizational specifications except roles have been generated because agents cannot communicate. For the KAZ environment, we can notice two distinct roles: archers tend to move away from zombies, while knights tend to approach them. For the PPY environment, we can observe the output specifications indicate authority links between the leader predator and the simple predators to enable collective strategies for circling prey. Finally, the CYB environment shows communications between blue agents are indeed understood as communication links that enable isolating infiltrated drones or trying to fix and alert recently suspected drones.

For the CYB environment, we developed our custom MAS via a simple hand-crafted decision tree as preconized in AOMEA in light of the organizational specifications we curated by removing noisy results. Our approach did not suggest general roles but relevant strategy patterns have been identified. For instance, regarding links between agents' roles, we noticed that the agents sending messages frequently seem to be spotted as suspected by their neighbors. In addition, a cyber-defender agent in the communication radius of a suspected drone tends to switch off its communication and reactivate afterward. Even though these insights are few, the mean score we got with our curated MAS is about -2000 which is indeed close to the top 5 scores. This shows AOMEA to be indeed applicable to the Cyberdefense context additionally bringing safety guarantees.

\section{Conclusion}




MAS methodological works rely on the designer's knowledge to design a suited MAS organization but do not provide automatic or assisted ways to determine relevant organizational mechanisms.
MARL techniques have been successfully applied to train agents automatically to reach the given goal without explicit characterization of emergent collective strategies.
AOMEA's originality is to augment a MARL process with an explicit organizational model towards a methodological purpose to address these issues. We first exposed how AOMEA is intended to be used in MAS engineering as an additional tool to assist in the design process.
Then, we explained the AOMEA's theoretical core with links between Dec-POMDP and the $\mathcal{M}OISE^+$ through the \emph{PRAHOM} process.
Furthermore, we implemented the \emph{PRAHOM PettingZoo wrapper} as a Proof of Concept for practically applying AOMEA and we showed it enables getting some organizational specifications that satisfy the design constraints and allow achieving the given goal.
Finally, we applied our approach in four PettingZoo environments to assess the impact on and after training. The obtained performance results show to be comparable to known ones showing our approach to be viable.

Even though \emph{PRAHOM} is agnostic of the MARL algorithm because it uses agents' histories to infer organizational specifications, reconstructing agents' collective behaviors a posteriori may be difficult. Indeed, a major perspective for improving \emph{PRAHOM} is to go further with supervised and non-supervised learning techniques in addition to empirical statistical approaches for identifying valuable organizational specifications from joint-histories. Moreover, it is worth investigating recent works in MARL techniques such as hierarchical learning because they already seek to characterize emergent strategies throughout learning.

%
%
%
\section*{References}

\bibliographystyle{splncs04}

\bibliography{references.bib}

\end{document}